\documentclass[conference]{IEEEtran}

\ifCLASSINFOpdf
\else
\fi

\usepackage{amsmath}

\usepackage{amssymb}

\usepackage{booktabs}
\usepackage{multirow}
\usepackage{multicol}

\usepackage{graphicx}
\usepackage{color}
\usepackage{float}
\restylefloat{table}
\usepackage{placeins}
\usepackage{epstopdf}

\newcommand\pN{\mathcal{N}}
\newcommand\supinf{\underset{H_0}{\overset{H_1}{\gtrless}}}
 \usepackage{bbold}
\usepackage[T1]{fontenc}
\pdfmapfile{=fullembed.map}

\begin{document}

\title{Sparsity-based Cholesky Factorization and Its Application to Hyperspectral Anomaly Detection}

\author{\IEEEauthorblockN{Ahmad W. Bitar\IEEEauthorrefmark{1}, Jean-Philippe Ovarlez\IEEEauthorrefmark{1}\IEEEauthorrefmark{2} and Loong-Fah Cheong\IEEEauthorrefmark{3}}
\IEEEauthorblockA{\IEEEauthorrefmark{1}SONDRA/CentraleSup\'elec, Plateau du Moulon, 3 rue Joliot-Curie, F-91190 Gif-sur-Yvette, France\\
}
\IEEEauthorblockA{\IEEEauthorrefmark{2,1}ONERA, DEMR/TSI, Chemin de la Huni\`ere, 91120 Palaiseau, France\\ 
}
\IEEEauthorblockA{\IEEEauthorrefmark{3}National University of Singapore (NUS), Singapore, Singapore\\
}
}

\maketitle

\begin{abstract}
Estimating large covariance matrices has been a longstanding important problem in many applications and has attracted increased attention over several decades. This paper deals with two methods based on pre-existing works to impose sparsity on the covariance matrix via its unit lower triangular matrix (aka Cholesky factor) $\mathbf{T}$. The first method serves to estimate the entries of $\mathbf{T}$ using the Ordinary Least Squares (OLS), then imposes sparsity by exploiting some generalized thresholding techniques such as Soft and Smoothly Clipped Absolute Deviation (SCAD). The second method directly estimates a sparse version of $\mathbf{T}$ by penalizing the negative normal log-likelihood with $L_1$ and SCAD penalty functions.
The resulting covariance estimators are always guaranteed to be positive definite. 
Some Monte-Carlo simulations as well as experimental data demonstrate the effectiveness of our estimators for hyperspectral anomaly detection using the Kelly anomaly detector.
\end{abstract}
$\bf Keywords$-- Hyperspectral anomaly detection, covariance matrix, sparsity, Cholesky factor.

\IEEEpeerreviewmaketitle

%%%%%%%%%%%%%%% INTRODUCTION %%%%%%%%%%%%%%%%%%%%%%%%%%%%%%%%%%%%%%%%%%%%%%%%%%%%%%%%%%%%

\section{Introduction}
An airborne hyperspectral imaging sensor is capable of simultaneously acquiring the same spatial scene in a contiguous and multiple narrow spectral wavelength (color) bands \cite{Shaw02, manolakis2003hyperspectral, manolakis_lockwood_cooley_2016}. When all the spectral bands are stacked together, the resulting hyperspectral image (HSI) is a three dimensional data cube; each pixel in the HSI is a $p$-dimensional vector, $\mathbf{x} = [x_1, \, \cdots, \, x_p]^T \in \mathbb{R}^p$, where $p$ designates the total number of spectral bands. 
With the rich information afforded by the high spectral dimensionality, hyperspectral imagery has found many applications in various fields such as agriculture \cite{Patel2001, Datt2003}, mineralogy \cite{Lehmann2001}, military \cite{manolakis2002detection, Stein02, 4939406}, and in particular, target detection \cite{Shaw02, manolakis2003hyperspectral, Manolakis14, Manolakis09, manolakis2002detection, Frontera13, Frontera14}.
In many situations of practical interest, we do not have sufficient a priori information to specify the statistics of the target class. More precisely, the target's spectra is not provided to the user. This unknown target is referred as \guillemotleft~anomaly \guillemotright~\cite{matteoli2010tutorial} having a very different spectra from the background (e.g.,  a ship at sea).

Different Gaussian-based anomaly detectors have been proposed in the literature \cite{1990ITASSRXD, 4104190, 1020263, ff1993, 7739987}. The detection performance of these detectors mainly depend on the true unknown covariance matrix (of the background surrounding the test pixel) whose entries have to be carefully estimated specially in large dimensions. Due to the fact that in hyperspectral imagery, the number of covariance matrix parameters to estimate grows with the square of the spectral dimension, it becomes impractical to use traditional covariance estimators where the target detection performance can deteriorate significantly. 
Many a time, the researchers assume that compounding the large dimensionality problem can be alleviated by leveraging on the assumption that the true unknown covariance matrix is sparse, namely, many entries are zero.

This paper outlines two simple methods based on pre-existing works in order to impose sparsity on the covariance matrix via its Cholesky factor $\mathbf{T}$. The first method imposes sparsity by exploiting thresholding operators such as Soft and SCAD on the OLS estimate of $\mathbf{T}$. The second method directly estimates a sparse version of $\mathbf{T}$ by penalizing the negative normal log-likelihood with $L_1$ and SCAD penalty functions.

{\it Summary of Main Notations:} Throughout this paper, we depict vectors in lowercase boldface letters and matrices in uppercase boldface letters. The notation $(.)^T$ stands for the transpose, while $|.|$, $(.)^{-1}$, $(.)^{'}$, $\det(.)$, and $\mathbb{1}$ are the absolute value, the inverse, the derivative, the determinant, and indicator function, respectively. For any $z \in \mathbb{R}$, we define $sign(z)=1$ if $z > 0$, $sign(z)=0$ if $z = 0$ and $sign(z)=-1$ if $z < 0$.

\section{Background and System Overview}
Suppose that we observe a sample of $n$ independent and identically distributed $p$-random vectors, $\{\mathbf{x}_i\}_{i\in[1,\,n]}$, each follows a multivariate Gaussian distribution with zero mean and unknown covariance matrix $\boldsymbol{\Sigma} = [\sigma_{g,l}]_{p \times p}$.
The first traditional estimator we consider in this paper is the Sample Covariance Matrix (SCM), defined as ${\hat{\boldsymbol{\Sigma}}_{SCM}} = [\hat{\sigma}_{g,l}]_{p \times p} =  \displaystyle \frac{1}{n} \sum\limits_{i=1}^n \mathbf{x}_i \, \mathbf{x}_i^T$.
\\
In order to address the positivity definiteness constraint problem of ${\hat{\boldsymbol{\Sigma}}_{SCM}}$, Pourahmadi \cite{Pourahmadi99} has modeled the covariance matrices via linear regressions. This is done by  letting $\hat{\mathbf{x}} =[\hat{x}_1, \, \ldots, \, \hat{x}_p]^T \in \mathbb{R}^{p}$, and consider each element $\hat{x}_t$, $t \in [1, \, p]$, as the linear least squares predictor of $x_t$ based on its $t-1$ predecessors $\{x_j\}_{j\in[1,\,t-1]}$. In particular, for $t \in [1, \ p]$, let
\begin{equation}
\hat{x}_t = \sum\limits_{j=1}^{t-1} C_{t,j}\, x_j , \hspace{0.5cm} \mathbf{T}\boldsymbol{\Sigma}\mathbf{T}^T = \mathbf{D} \,.
\end{equation}
where $\mathbf{T}$ is a unit lower triangular matrix with $-C_{t,j}$ in the $(t,j)$th position for $t \in [2, \, p]$ and $j \in[1, \, t-1]$, and $\mathbf{D}$ is a diagonal matrix with $\theta_t^2 = \mathrm{var}(\epsilon_t)$ as its diagonal entries, where $\epsilon_t = x_t - \hat{x}_t$ is the prediction error for $t \in [1, \ p]$. Note that for $t=1$, let $\hat{x}_1 = E(x_1) = 0$, and hence, $\mathrm{var}(\epsilon_1) = \theta_1^2 = E\left[\left(x_1\right)^2\right]$.
Given a sample $\{\mathbf{x}_i\}_{i\in[1,\,n]}$, with $n>p$, a natural estimate of $\mathbf{T}$ and $ \mathbf{D}$, denoted as $\hat{\mathbf{T}}_{OLS}$ and $\hat{\mathbf{D}}_{OLS}$ in this paper, is simply done by plugging in the OLS estimates of the regression coefficients and residual variances in (1), respectively. In this paper, we shall denote the second traditional estimator by $\hat{\boldsymbol{\Sigma}}_{OLS} = \hat{\mathbf{T}}^{-1}_{OLS} \, \hat{\mathbf{D}}_{OLS} \, \hat{\mathbf{T}}^{-T}_{OLS}$.

Obviously, when the spectral dimension $p$ is considered large compared to the number of observed data $n$, both ${\hat{\boldsymbol{\Sigma}}_{SCM}}$ and ${\hat{\boldsymbol{\Sigma}}_{OLS}}$ face difficulties in estimating $\boldsymbol{\Sigma}$ without an extreme amount of errors. Realizing the challenges brought by the high dimensionality, researchers have thus circumvent these challenges by proposing various regularization techniques to consistently estimate $\boldsymbol{\Sigma}$ based on the assumption that the covariance matrix is sparse.
Recently, Bickel et al. \cite{Bickel08} proposed a banded version of ${\hat{\boldsymbol{\Sigma}}_{SCM}}$, denoted as $B_m (\hat{\boldsymbol{\Sigma}}_{SCM})$ in this paper, with $B_m (\hat{\boldsymbol{\Sigma}}_{SCM}) = [\hat{\sigma}_{g,l}\,\mathbb{1}(|g-l| \leq m)]$, where $0 \leq m<p$ is the banding parameter. However, this kind of regularization does not always guarantee positive definiteness.
\\
In \cite{Rothman09}, a class of generalized thresholding operators applied on the off-diagonal entries of ${\hat{\boldsymbol{\Sigma}}_{SCM}}$ have been discussed. These operators combine shrinkage with thresholding and have the advantage to estimate the true zeros as zeros with high probability.  These operators (e.g., Soft and SCAD), though simple, do not always guarantee positive definiteness of the thresholded version of ${\hat{\boldsymbol{\Sigma}}_{SCM}}$.
In \cite{Cao09}, the covariance matrix is constrained to have an eigen decomposition which can be represented as a sparse matrix transform (SMT) that decomposes the eigen-decomposition into a product of very sparse transformations. The resulting estimator, denoted as $\hat{\boldsymbol{\Sigma}}_{SMT}$ in this paper, is always guaranteed to be positive definite.
\\
In addition to the above review, some other works have attempted to enforce sparsity on the covariance matrix via its Cholesky factor $\mathbf{T}$. Hence, yielding sparse covariance estimators that are always guaranteed to be positive definite. For example, in \cite{WhoAndPourhmadiSmoothing}, Pourahmadi et al. proposed to smooth the first few subdiagonals of $\hat{\mathbf{T}}_{OLS}$ and set to zero the remaining subdiagonals. In \cite{Huang06}, Huang {\it et al.} 
proposed to directly estimate a sparse version of $\mathbf{T}$ by penalizing the negative normal log-likelihood with a $L_1$-norm penalty function. Hence, allowing the zeros to be irregularly placed in the Cholesky factor. This seems to be an advantage over the work in \cite{WhoAndPourhmadiSmoothing}.

We put forth two simple methods for imposing sparsity on the covariance matrix via its Cholesky factor $\mathbf{T}$. The first method is based on the work in \cite{Rothman09}, but attempts to render $\hat{\boldsymbol{\Sigma}}_{OLS}$ sparse by thresholding its Cholesky factor $\hat{\mathbf{T}}_{OLS}$ using operators such as Soft and SCAD. The second method aims to generalize the work in \cite{Huang06} in order to be used for various penalty functions. 
The two methods allow the zeros to be irregularly placed in the Cholesky factor. 
\\
Clearly, in real world hyperspectral imagery, the true covariance model is not known, and hence, there is no prior information on its degree of sparsity.
However, enforcing sparsity on the covariance matrix seems to be a strong assumption, but can be critically important if the true covariance model for a given HSI is indeed sparse. That is, taking advantage of the possible sparsity in the estimation can potentially improve the target detection performance, as can be seen from the experimental results later. 
On the other hand, while the true covariance model may not be sparse (or not highly sparse), there should be no worse detection results than to those of the traditional estimators $\hat{\boldsymbol{\Sigma}}_{SCM}$ and $\hat{\boldsymbol{\Sigma}}_{OLS}$.

We evaluate our estimators for hyperspectral anomaly detection using the Kelly anomaly detector \cite{Kelly86}. More precisely, we first perform a thorough evaluation of our estimators on some Monte-Carlo simulations for three true covariance models of different sparsity levels. From our experiments in Subsection \ref{sec:sub3}, the detection results show that in trully sparse models, our estimators improve the detection significantly with respect to the traditional ones, and have competitive results with state-of-the-art \cite{Bickel08, Rothman09, Cao09}. When the true model is not sparse, we find that empirically our estimators still have no worse detection results than to those of $\hat{\boldsymbol{\Sigma}}_{SCM}$ and $\hat{\boldsymbol{\Sigma}}_{OLS}$.
\\
Next, in Subsection \ref{sec:sub4}, our estimators are evaluated on experimental data where a good target detection performances are obtained comparing to the traditional estimators and state-of-the-art.
In all the experiments later, we observe that $\hat{\boldsymbol{\Sigma}}_{OLS}$ achieves higher target detection results than to those of $\hat{\boldsymbol{\Sigma}}_{SCM}$.

%%%%%%%%%%%%%%%%%%%% MAIN CONTRIBUTIONS %%%%%%%%%%%%%%%%%%%%%%%%%%%%%%%%%%%%%%%%%%%%%%

\section{Main Contributions}
Before describing the two methods, we want to recall the definition for $\hat{\boldsymbol{\Sigma}}_{OLS}$. Given a sample $\{\mathbf{x}_i\}_{i\in[1,\,n]}$, we have:
\begin{equation}
\label{eq:eq2}
x_{i,t} = \sum\limits_{j=1}^{t-1} C_{t,j} \, x_{i,j} + \epsilon_{i,t} \hspace{0.5cm} t \in [2, \, p], \hspace{0.5cm}  i\in[1, \, n] .
\end{equation}
By writing \eqref{eq:eq2} in vector-matrix form for any $t \in [2, \,p]$, one obtains the simple linear regression model:
\begin{equation}
\label{eq:eq3}
\mathbf{y}_t = \mathbf{A}_{n,t} \, \boldsymbol{\beta}_t + \mathbf{e}_t \,,
\end{equation}
where $\mathbf{y}_t  = [x_{1,t}, \, \cdots, \, x_{n,t}]^T\in \mathbb{R}^{n}$, $\mathbf{A}_{n,t} =[x_{i,j}]_{n \times (t-1)}$, $\boldsymbol{\beta}_t  = [C_{t,1}, \, \cdots, \, C_{t,t-1}]^T \in \mathbb{R}^{(t-1)}$, and $\mathbf{e}_t  = [\epsilon_{1,t}, \, \cdots, \, \epsilon_{n,t}]^T \in \mathbb{R}^n$.
\\
When $n>p$, the OLS estimate of $\boldsymbol{\beta}_t$ and the corresponding residual variance are plugged in $\mathbf{T}$ and $\mathbf{D}$ for each $t\in[2, \, p]$, respectively. At the end, one obtains the estimator $\hat{\boldsymbol{\Sigma}}_{OLS} = \hat{\mathbf{T}}^{-1}_{OLS} \, \hat{\mathbf{D}}_{OLS} \, \hat{\mathbf{T}}^{-T}_{OLS}$. Note that $\hat{\mathbf{T}}_{OLS}$ has -$\hat{C}^{OLS}_{t,j}$ in the $(t,j)$th position for $t \in [2, \, p]$ and $j \in [1, \, t-1]$.
 
\subsection{Generalized thresholding based Cholesky Factor}
\label{sec:sub1}
For any $0 \leq \lambda \leq 1$, we define a matrix thresholding operator $Th(.)$ and denote by $Th(\hat{\mathbf{T}}_{OLS}) = [Th(-\hat{C}_{t,j}^{OLS})]_{p \times p}$ the matrix resulting from applying a specific thresholding operator $Th(.) \in \{$Soft, SCAD$\}$ to each element of the matrix $\hat{\mathbf{T}}^{OLS}$ for $t \in [2, \, p]$ and $j \in [1, \, t-1]$.
\\
We consider the following minimization problem: 
\begin{equation}
\label{eq:eq4}
\footnotesize
%\begin{split}
%Th(\hat{\mathbf{T}}_{OLS}) = \underset{\mathbf{T}} {\mathrm{argmin}} \frac{1}{2} ||\mathbf{T} - \mathbf{\hat{T}}_{OLS}||^2_F + \sum \limits_{t=2}^p \sum\limits_{j=1}^{t-1}p_{\lambda}\{|C_{t,j}|\}
%\\
Th(\hat{\mathbf{T}}_{OLS}) =  \underset{\mathbf{T}} {\mathrm{argmin}} \sum\limits_{t=2}^p \sum\limits_{j=1}^{t-1} \bigg\{\frac{1}{2}(\hat{C}_{t,j}^{OLS} - C_{t,j})^2 + p_{\lambda}\{|C_{t,j}|\}\bigg\}
%\end{split}
\end{equation}
where $p_{\lambda} \in \{p^{L_1}_{\lambda}, p^{SCAD}_{\lambda, a>2}\}$.
We have 
$p_{\lambda}^{L_1}(\mathopen|C_{t,j}\mathclose|)$ = $\lambda \mathopen|C_{t,j}\mathclose|$, and $p_{\lambda, a>2}^{SCAD}(\mathopen|C_{t,j}\mathclose|)$  = $\left\{ \begin{array}{ll} \lambda \mathopen|C_{t,j}\mathclose| & \mbox{if } \mathopen|C_{t,j}\mathclose| \leq \lambda \\ -\frac{\mathopen|C_{t,j}^2 \mathclose| - 2a\lambda \mathopen|C_{t,j}\mathclose| + \lambda^2}{2(a-1)} & \mbox{if } \lambda<\mathopen|C_{t,j}\mathclose|\le a\lambda \\ \frac{(a+1)\lambda^2}{2}  & \mbox{if } \mathopen|C_{t,j}\mathclose|>a\lambda   \end{array} \right.$.
\\
\\
Solving \eqref{eq:eq4} with $p^{L_1}_{\lambda}$ and $p^{SCAD}_{\lambda, a>2}$, yields a closed-form Soft and SCAD thresholding rules, respectively {\cite{Rothman09}, \cite{Fan01}}. 
The value $a = 3.7$ was recommended by Fan and Li \cite{Fan01}. Despite the application here is different than in \cite{Fan01}, for simplicity, we use the same value throughout the paper.
\\
We shall designate the two thresholded matrices by $\hat{\mathbf{T}}_{Soft}$ and $\hat{\mathbf{T}}_{SCAD}$, that apply Soft and SCAD on $\hat{\mathbf{T}}_{OLS}$, respectively. 
We denote our first two estimators as:
\vspace{-0.3cm}
\[
 \boxed{\hat{\boldsymbol{\Sigma}}_{OLS}^{Soft} =  \hat{\mathbf{T}}^{-1}_{Soft} \, \hat{\mathbf{D}}_{OLS} \, \hat{\mathbf{T}}^{-T}_{Soft}}
 \]
\vspace{-0.2cm}
 \[
 \boxed{\hat{\boldsymbol{\Sigma}}_{OLS}^{SCAD} =  \hat{\mathbf{T}}^{-1}_{SCAD} \, \hat{\mathbf{D}}_{OLS} \hat{\mathbf{T}}^{-T}_{SCAD}}
 \]
\begin{table*}[t]
\begin{center}
\resizebox{18.15cm}{!}{
\begin{tabular}{ | l | l | l | | l | l | l | | l | l | l | | l | l | l | l | l | l |}
\hline
~~\bf{Models} & ~~~~~$\boldsymbol{\Sigma}$ & $\hat{\boldsymbol{\Sigma}}_{SCM}$ & $\hat{\boldsymbol{\Sigma}}_{OLS}$ & $\color{red}\hat{\boldsymbol{\Sigma}}_{OLS}^{Soft}$ & $\color{red}\hat{\boldsymbol{\Sigma}}_{OLS}^{SCAD}$ & $\color{red}\hat{\boldsymbol{\Sigma}}_{L_1}$ & $\color{red}\hat{\boldsymbol{\Sigma}}_{SCAD}$ & $\hat{\boldsymbol{\Sigma}}_{SMT}$ & $B_k (\hat{\boldsymbol{\Sigma}}_{SCM})$ & $\hat{\boldsymbol{\Sigma}}_{SCM}^{Soft}$ & $\hat{\boldsymbol{\Sigma}}_{SCM}^{SCAD}$ \\
\hline
\bf~~Model 1 &  ~~0.9541 & 0.7976 & 0.8331 & 0.9480 & 0.9480 & \bf{0.9509} & \bf{0.9509} & 0.9503 & ~~\bf{0.9509} & \bf{0.9509} & \bf{0.9509}\\
\hline
\bf~~Model 2 & ~~0.9540 & 0.7977 & 0.8361 & 0.9124 & 0.9124 & 0.9264 & 0.9264 & 0.9184 & ~~\bf{0.9478} & 0.9274 & 0.9270\\
\hline
\bf~~Model 3 & ~~0.9541 & 0.7978 & 0.8259 & 0.8169 &  0.8257 & 0.8236 & \bf{0.8261} & 0.7798 & ~~0.5321 & 0.5969 & 0.5781\\
\hline
\bf~~MUSE & Not known & 0.6277 & 0.6575 & 0.9620 & \bf{0.9643} & 0.8844 & 0.8844 & 0.7879 & ~~0.9277 & 0.7180 & 0.7180\\
\hline
\end{tabular}
}
\end{center}
~~{\footnotesize {\bf Table 1. }A List of Area Under Curve (AUC) values of our estimators $\hat{\boldsymbol{\Sigma}}_{OLS}^{Soft}$, $\hat{\boldsymbol{\Sigma}}_{OLS}^{SCAD}$, $\hat{\boldsymbol{\Sigma}}_{L_1}$, $\hat{\boldsymbol{\Sigma}}_{SCAD}$} when compared to some others.
\end{table*}
Note that in \cite{Rothman09}, the authors have demonstrated that for a non sparse true covariance model, the covariance matrix does not suffer any degradation when thresholding is applied to the off-diagonal entries of $\hat{\boldsymbol{\Sigma}}_{SCM}$. However, this is not the case for the target detection problem where the inverse covariance is used; we found that, and in contrast to our estimators, the scheme in \cite{Rothman09} has a deleterious effect on the detection performance when compared to those of $\hat{\boldsymbol{\Sigma}}_{SCM}$ and $\hat{\boldsymbol{\Sigma}}_{OLS}$.

%%%%%%%%%%%%%% Q generalization of the estimator  %%%%%%%%%%%%%%%%%%%%%%%%%%%%%%%%%%%%%%%%%%%%%%%%%%%%%%

\subsection{A generalization of the estimator in \cite{Huang06}}
\label{sec:sub2}
We present the same concept in \cite{Huang06}, but by modifying the procedure by which the entries of $\mathbf{T}$ have been estimated. 
Note that $\det(\mathbf{T}) = 1$ and $\boldsymbol{\Sigma} = \mathbf{T}^{-1} \, \mathbf{D} \, \mathbf{T}^{-T}$. It follows that $\det(\boldsymbol{\Sigma}) = \det(\mathbf{D})= \prod \limits_{t=1}^{p} \theta_t^2$. Hence, the negative normal log-likelihood of $\mathbf{X} = [\mathbf{x}_1, \, \cdots, \, \mathbf{x}_n] \in \mathbb{R}^{p \times n}$, ignoring an irrelevant constant, satisfies: 
\\
$\Lambda = -2\log(L(\boldsymbol{\Sigma}, \mathbf{x}_1, \cdots, \mathbf{x}_n)) = n\log(\det(\mathbf{D})) + \mathbf{X}^{T} \, (\mathbf{T}^{T} \, \mathbf{D}^{-1} \, \mathbf{T}) \, \mathbf{X} =  n\log(\det(\mathbf{D})) + (\mathbf{T\, X})^T \, \mathbf{D}^{-1} (\mathbf{T \, X}) = n\sum\limits_{t=1}^{p} \log \theta_t^2 + \sum\limits_{t=1}^{p}\sum\limits_{i=1}^{n} \epsilon_{i,t}^2/\theta_{t}^2$.
By adding a penalty function $\sum\limits_{t=2}^{p}\sum\limits_{j=1}^{t-1} p_{\alpha} \{|C_{t,j}|\}$  to $\Lambda$, where $p_{\alpha} \in \{ p_{\alpha}^{L_1}, \, p_{\alpha,a>2}^{SCAD} \}$ (see subsection III. A) with $\alpha \in [0, \infty)$, we have:
\begin{equation}
\label{eq:eq5}
\footnotesize
n \log \theta_1^2 + \sum\limits_{i=1}^n \frac{\epsilon_{i,1}^2}{\theta_1^2} + \sum\limits_{t=2}^p \bigg(n \log~\theta_t^2 + \sum\limits_{i=1}^n \frac{\epsilon_{i,t}^{2}}{\theta_t^2}+ \sum\limits_{j=1}^{t-1} p_{\alpha} \{|C_{t,j}|\}\bigg)\,
\end{equation} 
Obviously, minimizing \eqref{eq:eq5} with respect to $\theta_1^2$ and $\theta_t^2$ gives the solutions $\hat\theta_1^2 = \frac{1}{n}\sum\limits_{i=1}^n \epsilon_{i,1}^2 = \frac{1}{n}\sum\limits_{i=1}^n x_{i,1}^2$ and $\hat\theta_t^2 = \frac{1}{n}\sum\limits_{i=1}^n \epsilon_{i,t}^2 = \frac{1}{n}\sum\limits_{i=1}^n (x_{i,t} - \sum\limits_{j=1}^{t-1} C_{t,j} x_{i,j})^2$, respectively. 
\\
It remains to estimate the entries of $\mathbf{T}$ by minimizing \eqref{eq:eq5} with respect to $\boldsymbol{\beta}_t$.
 From equation \eqref{eq:eq2} and \eqref{eq:eq3}, the minimization problem to solve for each $t \in [2, \, p]$ is:
\begin{equation}
\label{eq:eq6}
\small
\begin{split}
 \hat{\boldsymbol{\beta}}_t = \underset{\boldsymbol{\beta}_t} {\mathrm{argmin}} \sum\limits_{i=1}^n \frac{\epsilon_{i,t}^{2}}{\theta_t^2}+ \sum\limits_{j=1}^{t-1} p_{\alpha} \{|C_{t,j}|\} \\
      = \underset{\boldsymbol{\beta}_t} {\mathrm{argmin}} \frac{1}{\theta_t^2} \sum\limits_{i=1}^n \left(x_{i,t} - \sum\limits_{j=1}^{t-1} C_{t,j} x_{i,j}\, \right)^2+ \sum \limits_{j=1}^{t-1} p_{\alpha} \{ |C_{t,j}| \}\, \\
      = \underset{\boldsymbol{\beta}_{t}} {\mathrm{argmin}} \frac{1}{\theta_t^2} || \mathbf{y}_t - \mathbf{A}_{n,t} \boldsymbol{\beta}_t||_F^2    + \sum \limits_{j=1}^{t-1} p_{\alpha} \{ |C_{t,j}|\}\
\end{split}
\end{equation}
By denoting $l(\boldsymbol{\beta}_t) = \frac{1}{\theta_t^2} || \mathbf{y}_t - \mathbf{A}_{n,t} \boldsymbol{\beta}_t||_F^2$ and $r(\boldsymbol{\beta}_t) = \sum \limits_{j=1}^{t-1} p_{\alpha} \{|C_{t,j}|\} = \sum\limits_{j=1}^{t-1}r_j(C_{t,j})$,  we solve $\boldsymbol{\beta}_t$ iteratively using the General Iterative Shrinkage and Thresholding (GIST) algorithm \cite{Gong13a}:
\begin{equation}
\label{eq:eq7}
\footnotesize
\begin{split}
\hat{\boldsymbol{\beta}}_t^{(k+1)} = \underset{\boldsymbol{\beta}_{t}}{\mathrm{argmin}}~ l(\boldsymbol{\beta}_t^{(k)}) + r(\boldsymbol{\beta}_t) + (\nabla{l(\boldsymbol{\beta}_t^{(k)}}))^T (\boldsymbol{\beta}_t - \boldsymbol{\beta}_t^{(k)})\\
 + \frac{w^{(k)}}{2} ||\boldsymbol{\beta}_t - \boldsymbol{\beta}_t^{(k)}||^2
\\  = \underset{\boldsymbol{\beta}_{t}}{\mathrm{argmin}}~0.5 ||\boldsymbol{\beta}_t  - \mathbf{u}_t^{(k)}||^2 + \frac{1}{w^{(k)}} r(\boldsymbol{\beta}_t)
\end{split}
\end{equation}
where $\mathbf{u}_t^{(k)} = \boldsymbol{\beta}_t^{(k)} - \nabla l(\boldsymbol{\beta}_t^{(k)})/w^{(k)}$, and $w^{(k)}$ is the step size initialized using the Barzilai-Browein rule \cite{Barzilai88}.
\\
By decomposing \eqref{eq:eq7} into independent (t-1) univariate optimization problems, we have for $j = 1, \cdots, t-1$:
\begin{equation}
\label{eq:eq8}
\small
 C_{t,j}^{(k+1)} = \underset{C_{t,j}}{\mathrm{argmin}}~ 0.5 ||C_{t,j}  - u_{t,j}^{(k)}||^2 + \frac{1}{w^{(k)}} r_j(C_{t,j})
\end{equation}
where $\mathbf{u}_t^{(k)} = [u^{(k)}_{t,1} , \cdots, u^{(k)}_{t,t-1}]^T \in \mathbb{R}^{(t-1)}$.
\\
By solving \eqref{eq:eq8} with the $L_1$-norm penalty, $p_{\alpha}^{L_1}$, we have the following closed form solution:
\begin{equation}
C^{(k+1)}_{t,j, (L_1)} = sign(u^{(k)}_{t,j}) \, \max(0, \, |u^{(k)}_{t,j}| - \alpha/w^{(k)})
\end{equation}
For the SCAD penalty function, $p_{\alpha,a>2}^{SCAD}$, we can observe that it contains three parts for three different conditions (see Subsection \ref{sec:sub1}). In this case, by recasting problem \eqref{eq:eq8} into three minimization sub-problems for each condition, and after solving them, one can obtain the following three sub-solutions $h_{t,j}^1$, $h_{t,j}^2$, and $h_{t,j}^3$, where: 
\\
\\
$h_{t,j}^{1} = sign(u_{t,j}^{(k)}) \min(\alpha, \max(0, |u_{t,j}^{(k)}| - \alpha/w^{(k)}))$,
\\
$h_{t,j}^{2} = sign(u_{t,j}^{(k)}) \min(a\alpha, \max(\alpha, \frac{w^{(k)} |u_{t,j}^{(k)}| (a-1) - a\alpha }{w^{(k)} (a-2)}))$,
\\
$h_{t,j}^{3} = sign(u_{t,j}^{(k)}) \max(a\alpha, |u_{t,j}^{(k)}|)$.
\\
\\
Hence, we have the following closed form solution:
\begin{equation}
\small
\begin{split}
C^{(k+1)}_{t,j, (SCAD)} =  \underset{q_{t,j}}{\mathrm{argmin}}~0.5 (q_{t,j}  - u_{t,j}^{(k)})^2 + \frac{1}{w^{(k)}} r_j(q_{t,j})
\\
s.t. ~~~q_{t,j} \in \{ h_{t,j}^1, \, h_{t,j}^2, \, h_{t,j}^3\}
\end{split}
\end{equation}
\\
At the end, we denote our last two estimators as:
\[
 \boxed{\hat{\boldsymbol{\Sigma}}_{L_1} = \hat{\mathbf{T}}^{-1}_{L_1} \, \hat{\mathbf{D}} \, \hat{\mathbf{T}}^{-T}_{L_1}}
 \]
\vspace{-0.25cm}
 \[
 \boxed{\hat{\boldsymbol{\Sigma}}_{SCAD} = \hat{\mathbf{T}}^{-1}_{SCAD} \, \hat{\mathbf{D}} \, \hat{\mathbf{T}}^{-T}_{SCAD}}
 \]
where $ \hat{\mathbf{T}}_{L_1}$ and  $\hat{\mathbf{T}}_{SCAD}$ have respectively $-\hat{C}_{t,j,(L_1)}$ and $-\hat{C}_{t,j,(SCAD)}$ in the $(t,j)$th position for $t \in [2, \, p]$ and $j \in[1, \, t-1]$, whereas $\hat{\mathbf{D}}$ has the entries $(\hat{\theta}^{2}_1,\, \hat{\theta}^{2}_t)$ on its diagonal.
Note that in \cite{Huang06}, the authors have used the local quadratic approximation (LQA) of the $L_1$-norm in order to get a closed form solution for $\boldsymbol{\beta}_t$ in equation \eqref{eq:eq6}.
Our algorithm is now more general since after exploiting the GIST algorithm to solve \eqref{eq:eq6}, it can be easily extended to some other penalties such as SCAD \cite{Fan01}
%\cite{Fan01}
, Capped-L1 penalty \cite{Zhanggg, zhang2013, Gong12}, 
%\cite{Gong12},
 Log Sum Penalty\cite{Candes08}, Minimax Concave Penalty \cite{zhang2010aa} 
%\cite{Candes08},
etc. and they all have closed-form solutions \cite{Gong13a}. In this paper, we are only interested to the $L_1$ and SCAD penalty functions.

%%%%%%%%%%%%%% HYPERSPECTRAL ANOMALY DETECTION %%%%%%%%%%%%%%%%%%%%%%%%%%%%%%%%%%%%%%%%%%%%%%%%%%

\section{hyperspectral anomaly detection}
We first describe the Kelly anomaly detector \cite{Kelly86} used for the detection evaluation. Next, we present two subsections to gauge the detection performances of our estimators \{$\hat{\boldsymbol{\Sigma}}_{OLS}^{Soft}$,  $\hat{\boldsymbol{\Sigma}}_{OLS}^{SCAD}$, $\hat{\boldsymbol{\Sigma}}_{L_1}$, $\hat{\boldsymbol{\Sigma}}_{SCAD}$\} when compared to the traditional ones \{$\hat{\boldsymbol{\Sigma}}_{SCM}$, \, $\hat{\boldsymbol{\Sigma}}_{OLS}$\} and state-of-the-art: $\hat{\boldsymbol{\Sigma}}_{SMT}$ \cite{Cao09}, $B_k (\hat{\boldsymbol{\Sigma}}_{SCM})$ \cite{Bickel08}, and the two estimators that apply Soft and SCAD thresholding on the off-diagonal entries of $\hat{\boldsymbol{\Sigma}}_{SCM}$ in \cite{Rothman09}, and which will be denoted in the following experiments as $\hat{\boldsymbol{\Sigma}}_{SCM}^{Soft}$ and $\hat{\boldsymbol{\Sigma}}_{SCM}^{SCAD}$, respectively. Note that the tuning parameter $\lambda$ (in subsection \ref{sec:sub1}) and $\alpha$ (in subsection \ref{sec:sub2}) are chosen automatically using a 5-fold crossvalidated loglikelihood procedure (see Subsection 4.2 in \cite{Huang06} for details).

Suppose the following signal model:
\begin{equation} 
\small
\left\{
\begin{array}{l}
 H_0: \mathbf{x} = \mathbf{n}, ~~~~~~~~~~~~~~~\mathbf{x}_i = \mathbf{n}_i, ~~~~   i=1, \cdots, n\\
  H_1: \mathbf{x} = \gamma \,\mathbf{d}+ \mathbf{n}, ~~~~~~~~\mathbf{x}_i = \mathbf{n}_i, ~~~~   i=1, \cdots, n
\end{array}
\right.
\end{equation}
where $\mathbf{n}_1, \cdots, \mathbf{n}_n$ are $n$ i.i.d $p$-vectors, each follows a multivariate Normal distribution  $\pN(\mathbf{0}, \boldsymbol{\Sigma})$. $\mathbf{d}$ is an unknown steering vector and which denotes the presence of an anomalous signal with unknown amplitude $\gamma>0$.
After some calculation (refer to \cite{Kelly86} and both Subsection II. B and Remark II. 1 in \cite{Frontera16} for details), the Kelly anomaly detector is described as follows:
\begin{equation}
D_{KellyAD \hat{\boldsymbol{\Sigma}}}(\mathbf{x}) = \mathbf{x}^{T}\, \hat{\boldsymbol{\Sigma}}_{SCM}^{-1}\,\mathbf{x} \supinf \delta\, ,
\end{equation}
where $\delta$ is a prescribed threshold value.
In the following two subsections, the detection performances of the estimators, when are plugged in $D_{KellyAD, \hat{\boldsymbol{\Sigma}}}$ are evaluated by the Receiver Operating Characteristics (ROC) curves and their corresponding Area Under Curves (AUC) values. 

\begin{figure}[!tbp]
\minipage{0.16\textwidth}
  \includegraphics[width=\linewidth]{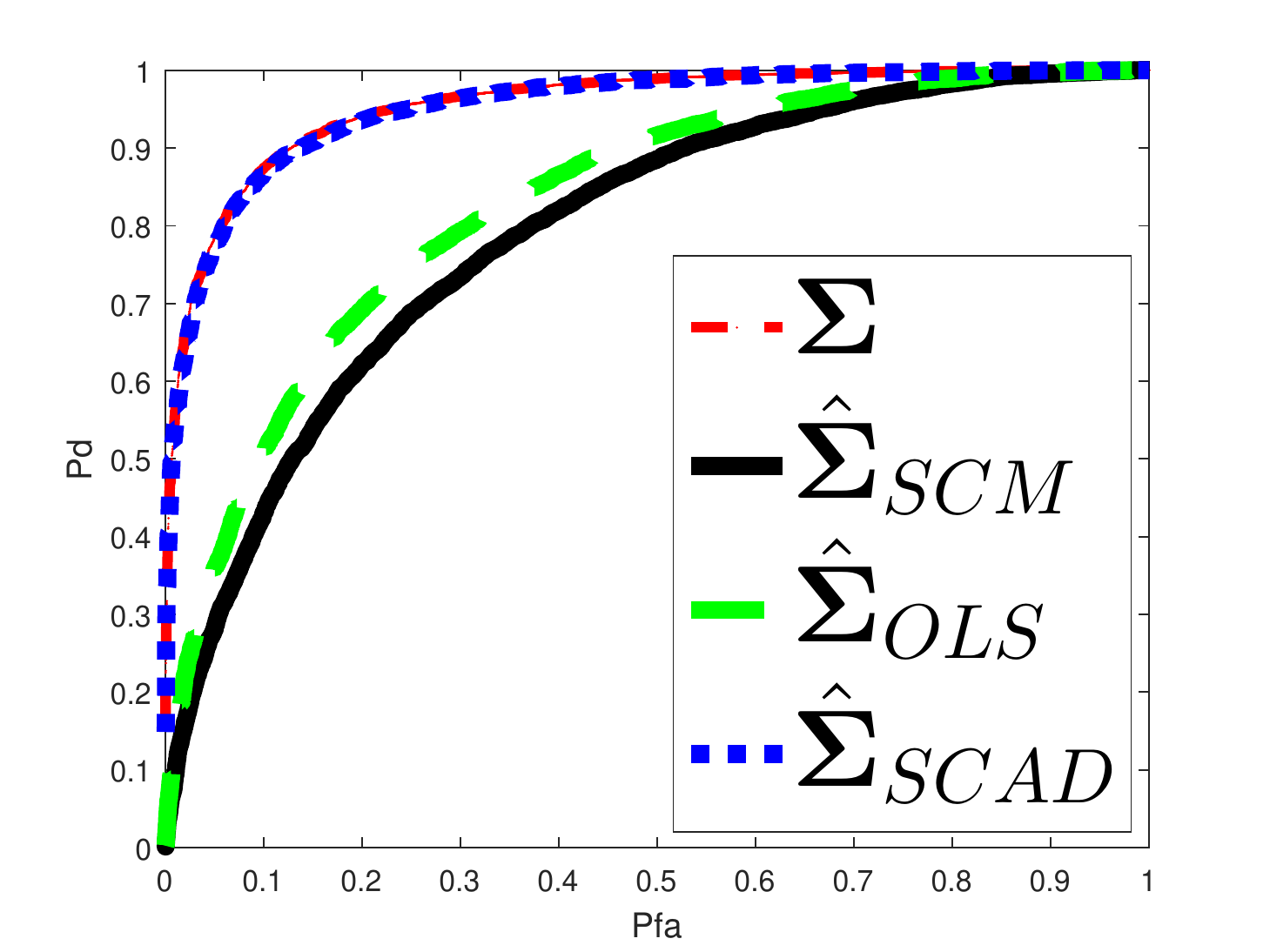}
{\begin{center}%
{\bf (a)}
\end{center}}
\endminipage\hfill
\minipage{0.16\textwidth}
  \includegraphics[width=\linewidth]{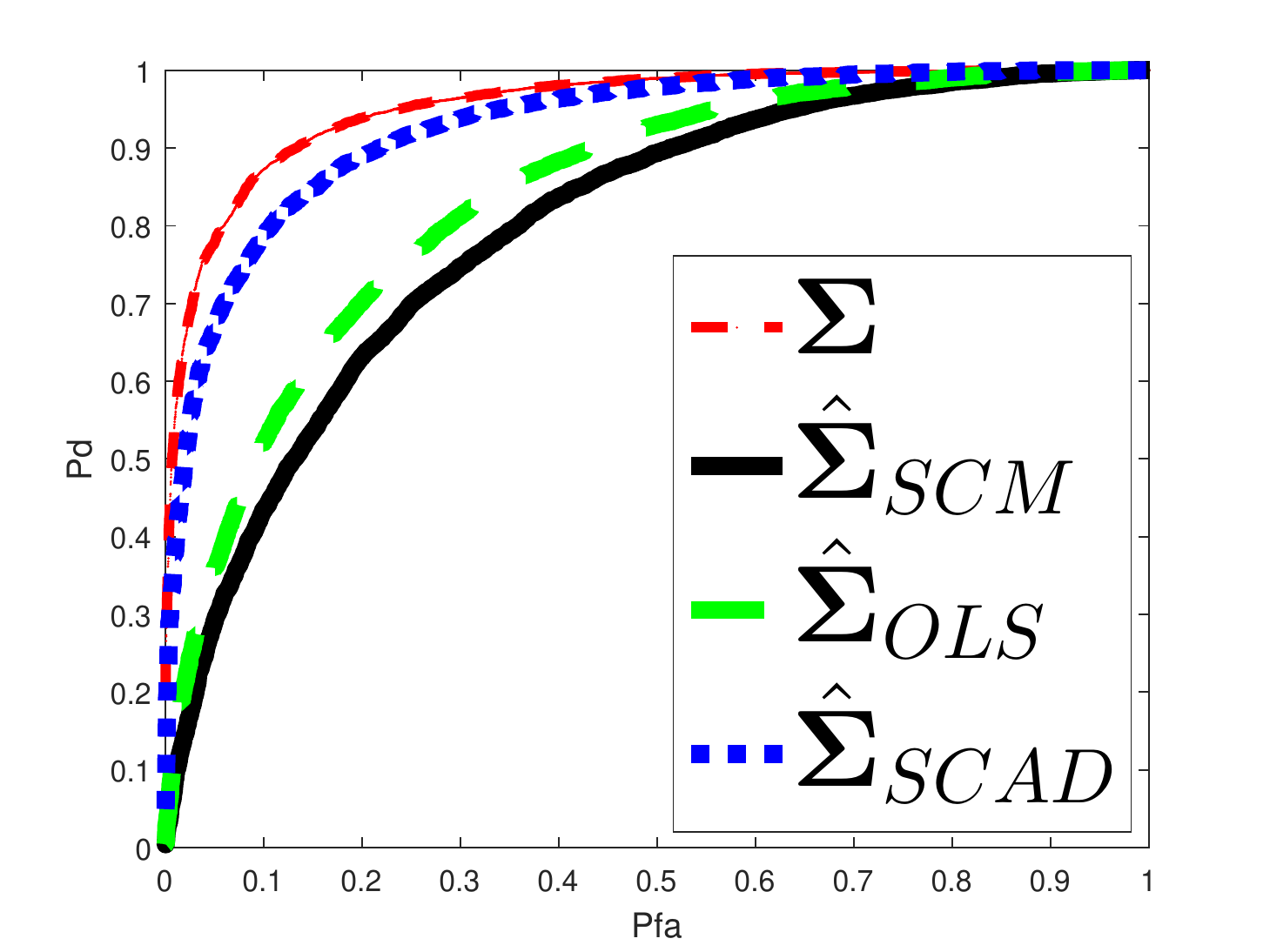}
{\begin{center}%
{\bf (b)}
\end{center}}
\endminipage\hfill
\centering
\minipage{0.16\textwidth}
  \includegraphics[width=\linewidth]{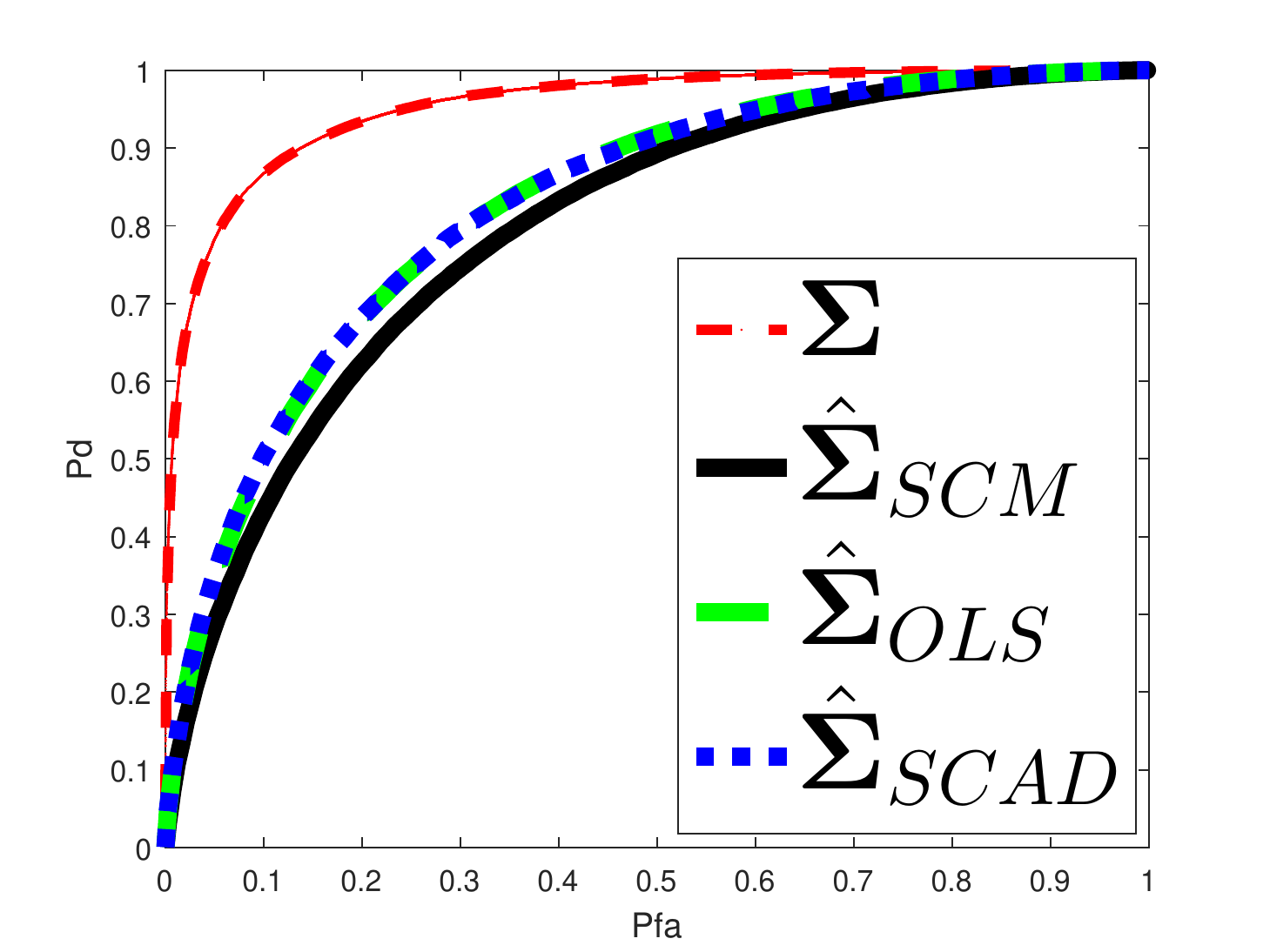}
{\begin{center}%
{\bf (c)}
\end{center}}
\endminipage
 \caption{ROC curves three Models. {\bf{(a)}} Model 1. {\bf{(b)}} Model 2. {\bf{(c)}} Model 3.}
\end{figure}
\begin{figure}[!htb]
\minipage{0.181\textwidth}
  \includegraphics[width=\linewidth]{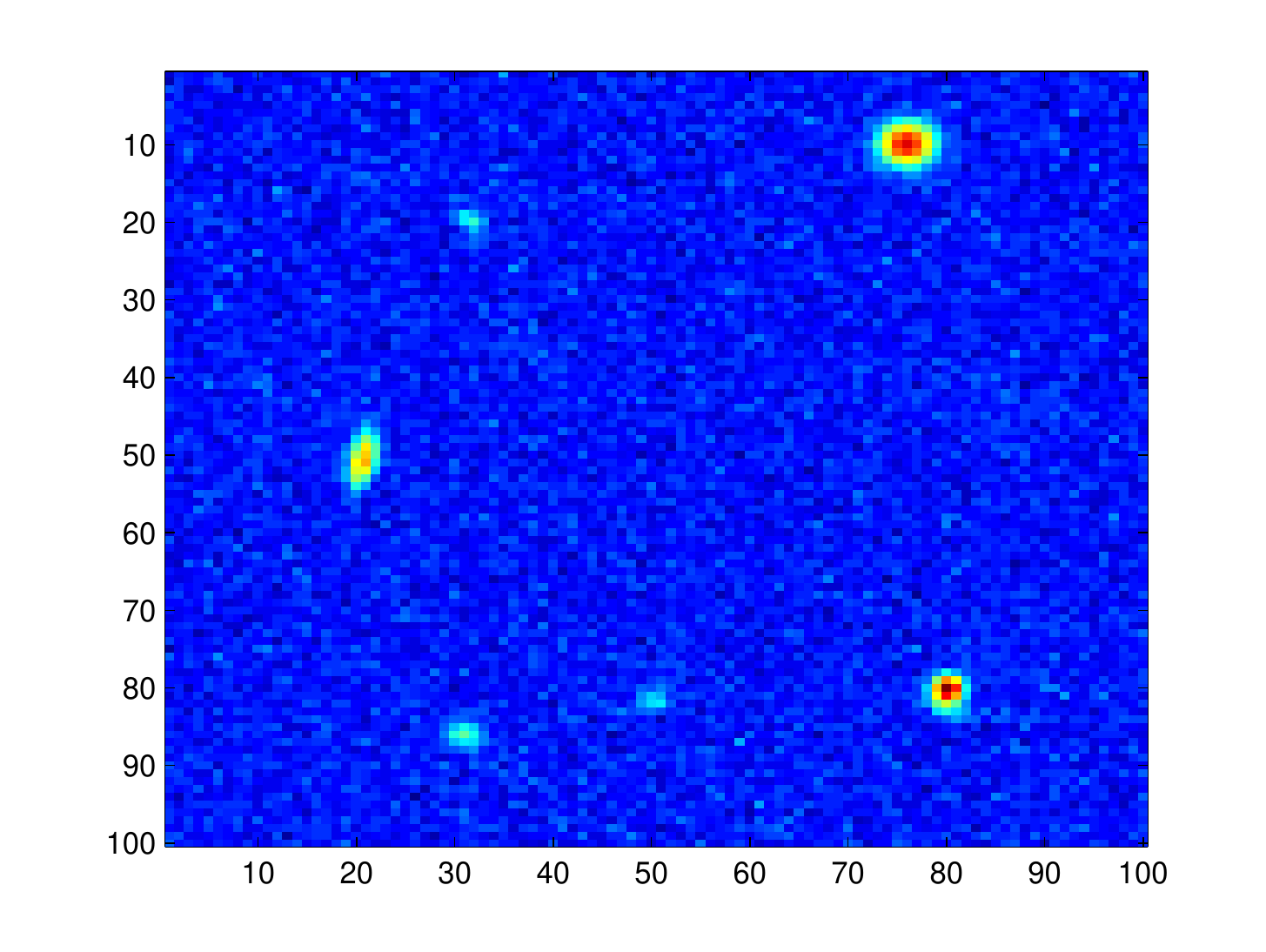}
{\begin{center}%
{\bf (a)}
\end{center}}
\endminipage\hfill
\minipage{0.19\textwidth}
  \includegraphics[width=\linewidth]{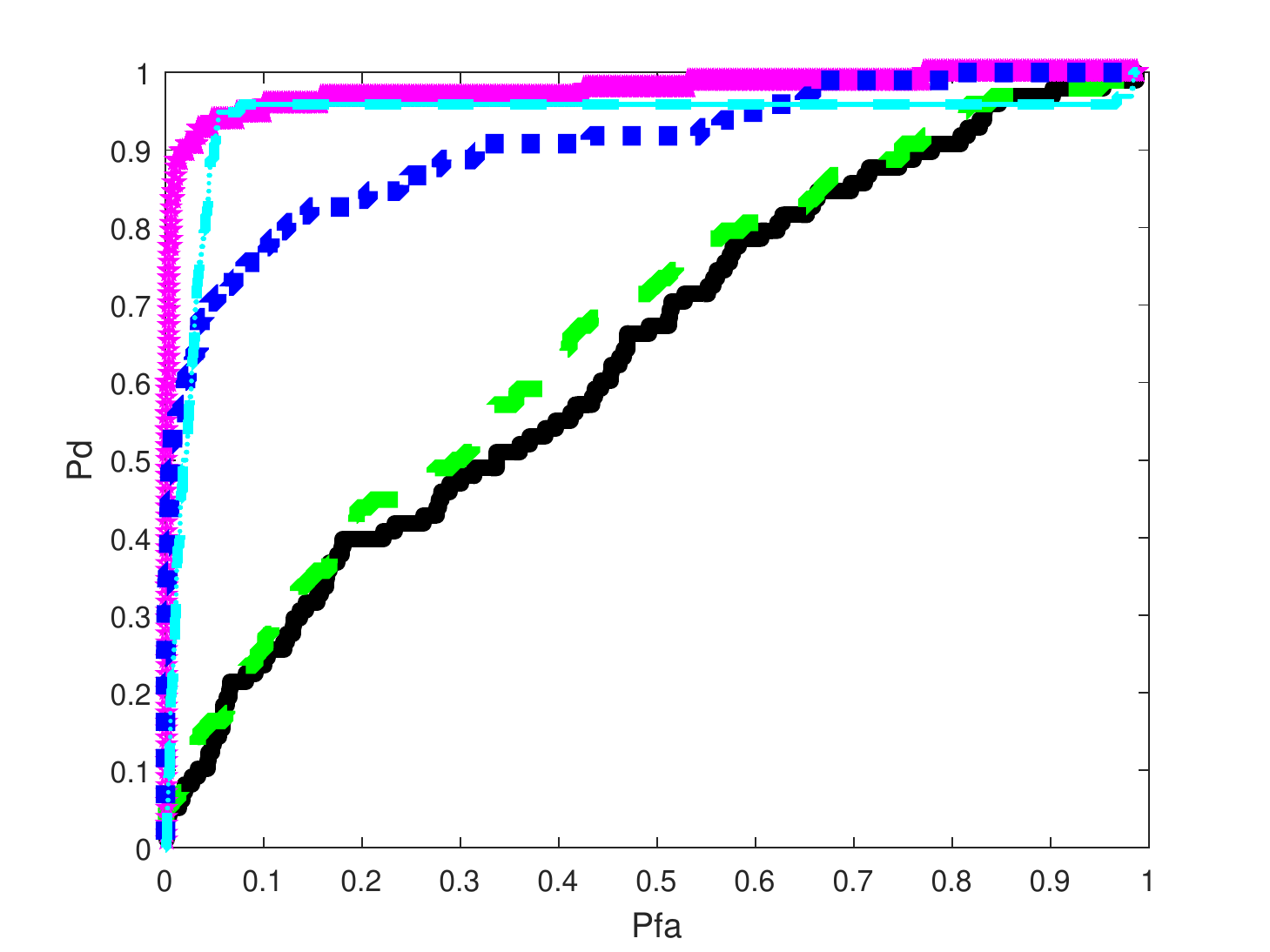}
{\begin{center}%
{\bf (b)}
\end{center}}
\endminipage\hfill
\minipage{.111\textwidth}
\includegraphics[width=\linewidth]{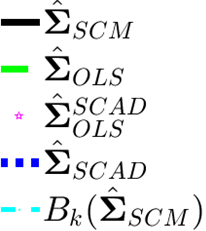}
{\begin{center}%
{\bf (c)}
\end{center}}
\endminipage
 \caption{{\bf{(a)}} MUSE HSI (average). {\bf{(b)}} ROC curves for MUSE. {\bf{(c)}} Legend.}
\end{figure}

%%%%%%%%%%%%%%%%%%%% Monte-Carlo SIMULATIONS %%%%%%%%%%%%%%%%%%%%%%%%%%%%%%%%%%%%%%%%%%%%%%%%%%

\subsection{Monte-Carlo simulations}
\label{sec:sub3}
The experiments are conducted on three covariance models: 
\begin{itemize}
\item Model 1: $\boldsymbol{\Sigma}$ = I, the identity matrix,
\item Model 2: the Autoregressive model order 1, AR(1), ${\boldsymbol{\Sigma}} = [\sigma_{gl}]_{p \times p}$, where $\sigma_{gl} = c^{|g-l|}$, for $c=0.3$,
\item Model 3: $\Sigma = [\sigma_{gl}]_{p \times p}$, where $\sigma_{gl} = (1-((|g-l|)/r))_{+}$, for $ r=p/2$: the triangular matrix.
\end{itemize}
Model 1 is very sparse and model 2 is approximately sparse. Model 3 with $r = p / 2$ is considered the least sparse \cite{Rothman09} among the three models we consider.
\\
The computations have been made through $10^5$ Monte-Carlo trials and the ROC curves are drawn for a signal to noise ratio equal to 15dB. We choose $n=80$ for covariance estimation under Gaussian assumption, and set $p$ = 60. The artificial anomaly we consider is a vector containing normally distributed pseudorandom numbers (to have fair results, the same vector is used for the three models). The ROC curves for Model 1, 2 and 3 are shown in Fig. 1{\bf (a)}, 1{\bf (b)} and 1{\bf (c)}, respectively, and their corresponding AUC values are presented in Table 1.
For a clear presentation of the figures, we only exhibit the ROC curves for $\boldsymbol{\Sigma}$, $\hat{\boldsymbol{\Sigma}}_{SCM}$, $\hat{\boldsymbol{\Sigma}}_{OLS}$, and $\hat{\boldsymbol{\Sigma}}_{SCAD}$.

The highest AUC values are shown in bold in Table 1.
For both Model 1 and 2, our estimators significantly improve the detection performances comparing to those of the traditional estimators ($\hat{\boldsymbol{\Sigma}}_{SCM}$, $\hat{\boldsymbol{\Sigma}}_{OLS}$), and have competitive detection results with state-of-the-art.
An important finding is that even for a non sparse covariance model (that is, Model 3) , our estimators do not show a harm on the detection when compared to those of $\hat{\boldsymbol{\Sigma}}_{SCM}$, $\hat{\boldsymbol{\Sigma}}_{OLS}$. Despite $\hat{\boldsymbol{\Sigma}}_{OLS}^{Soft}$, $\hat{\boldsymbol{\Sigma}}_{OLS}^{SCAD}$ and $\hat{\boldsymbol{\Sigma}}_{L_1}$ have slightly lower AUC values than for $\hat{\boldsymbol{\Sigma}}_{OLS}$, this is still a negligible degradation on the detection. Thus,
considering that $\hat{\boldsymbol{\Sigma}}_{OLS}^{Soft}$, $\hat{\boldsymbol{\Sigma}}_{OLS}^{SCAD}$ and $\hat{\boldsymbol{\Sigma}}_{L_1}$ have no worse detection results than to that of $\hat{\boldsymbol{\Sigma}}_{OLS}$ is still acceptable.

%%%%%%%%%%%%%%%%%%% APPLICATION ON EXPERIMENTAL DATA %%%%%%%%%%%%%%%%%%%%%%%%%%%%%%%%%%%%%%%%%%

\subsection{Application on experimental data}
\label{sec:sub4}
Our estimators are now evaluated for galaxy detection on the Multi Unit Spectroscopic Explorer (MUSE) data cube (see \cite{Muse}). It is a 100 $\times$ 100 image and consists of 3600 bands in wavelengths ranging from 465-930 nm. We used one band of each 60, so that 60 bands in total. 
Figure 2{\bf(a)} exhibits  the mean power in dB over the 60 bands. 
The covariance matrix is estimated using a sliding window of size $9 \times 9$, having $n = 80$ secondary data (after excluding only the test pixel). The mean has been removed from the given HSI. Figure 2{\bf(b)} exhibits the ROC curves \cite{Zhang15} of our estimators when compared to some others, and their AUC values are shown in Table 1. Note that the curves for $\hat{\boldsymbol{\Sigma}}_{SMT}$, $\hat{\boldsymbol{\Sigma}}_{SCM}^{Soft}$ and $\hat{\boldsymbol{\Sigma}}_{SCM}^{SCAD}$ are not drawn but their AUC values are shown in Table 1. 
\\
The estimators $\hat{\boldsymbol{\Sigma}}_{OLS}^{Soft}$, $\hat{\boldsymbol{\Sigma}}_{OLS}^{SCAD}$ achieve higher detection results than for all the others,
whereas both $\hat{\boldsymbol{\Sigma}}_{L_1}$ and $\hat{\boldsymbol{\Sigma}}_{SCAD}$ achieve only a lower AUC values than for $B_k (\hat{\boldsymbol{\Sigma}}_{SCM})$.

%%%%%%%%%%%%%%%% CONCLUSIN %%%%%%%%%%%%%%%%%%%%%%%%%%%%%%%%%%%%%%%%%%%%%%%%%%%%%%%%%%%%

\section{Conclusion}
Two methods are outlined to impose sparsity on the covariance matrix via its Cholesky factor $\mathbf{T}$. Some Monte-Carlo simulations as well as experimental data demonstrate the effectiveness (in terms of anomaly detection) of the two proposed methods using the Kelly anomaly detector.

%%%%%%%%%%%% BIBLIOGRAPHIE %%%%%%%%%%%%%%%%%%%%%%%%%%%%%%%%%%%%%%%%%%%%%%%%%%%%%%%%%%%%%%%

\bibliographystyle{IEEEtran}
\bibliography{biblio_these2}

\end{document}